\documentclass[%
 reprint,
superscriptaddress,
 amsmath,amssymb,
 aps,
prb,
]{revtex4-1}

\usepackage[utf8]{inputenc}
\usepackage{graphicx}
\usepackage{dcolumn}
\usepackage{bm}
\usepackage{hyperref}

\begin{document}

\title{A scalable method to find the shortest path in a graph with circuits of memristors}

\author{Alice Mizrahi}
\email{alicecmmizrahi@gmail.com}
\affiliation{National Institute of Standards and Technology, Gaithersburg, USA}
\affiliation{Maryland NanoCenter, University of Maryland, College Park, USA}

\author{Thomas Marsh}
\affiliation{National Institute of Standards and Technology, Gaithersburg, USA}

\author{Brian Hoskins}
\affiliation{National Institute of Standards and Technology, Gaithersburg, USA}

\author{M.D. Stiles}%
\affiliation{National Institute of Standards and Technology, Gaithersburg, USA}

\date{\today}

\begin{abstract}
Finding the shortest path in a graph has applications to a wide range of optimization problems. However, algorithmic methods scale with the size of the graph in terms of time and energy. We propose a method to solve the shortest path problem using circuits of nanodevices called memristors and validate it on graphs of different sizes and topologies. It is both valid for an experimentally derived memristor model and robust to device variability. The time and energy of the computation scale with the length of the shortest path rather than with the size of the graph, making this method particularly attractive for solving large graphs with small path lengths. 
\end{abstract}

\maketitle

\section{Introduction}
Optimization problems, such as sequence of decisions, resource allocation, or navigation, are omnipresent and important. They can be solved by being mapped to finding the shortest path between two nodes in a graph. Algorithms to do this exist, but the time and energy they consume scale with the size of the graph and thus become prohibitive for large systems.\cite{dijkstra_note_1959,ahuja_faster_1990,cherkassky_shortest_1996,thorup_undirected_1999,thorup_integer_2004,fredman_fibonacci_1984,williams_faster_2014} In this paper, we propose a method to solve the shortest path problem that scales consistently better than with the size of the graph. Our approach is inspired by biological systems that need to solve optimization problems to be energy efficient. For example, members of ant colonies work in parallel to find the shortest path to their food without supervision.\cite{bonabeau_inspiration_2000} Related algorithms, called ant colony optimization, take inspiration from this phenomenon to solve optimization problems in an approximate but efficient way.\cite{dorigo_ant_2006,parpinelli_data_2002,blum_ant_2005} More broadly, the field of swarm intelligence provides optimization algorithms inspired from animal populations. \cite{wahab_comprehensive_2015,ducatelle_principles_2010,poli_particle_2007} However, these algorithms are limited when they run on conventional hardware, which computes in a mostly sequential way. 

Pershin and Di Ventra proposed to solve the shortest path problem directly in hardware, using nanodevices called memristors, which have dynamics that provide reinforcement mechanisms similar to the ones at play in the ant colony optimization.\cite{pershin_solving_2011,pershin_self-organization_2013,pershin_memcomputing_2016,pajouhi_image_2018} Memristors are defined by having conductances that change when subjected to electrical current.\cite{strukov_missing_2008,pickett_switching_2009,menzel_origin_2011} As a voltage is applied to a network of memristors forming a graph, more current will flow through the shortest branch because that branch has the lowest resistance. This increased flow will cause an increase of conductance of the memristors on the shortest path, attracting even more current. When steady state is reached, the memristors on the shortest path have a much larger conductance than the ones on the longer paths, making it possible to electrically read out the shortest path. 

In this work, we build on Pershin and Di Ventra's idea and propose a modified method. The advantage of this method is that it does not require prior knowledge about the shortest path. We show that the time and energy consumed by this method scale with the length of the shortest path rather than with the size of the graph, which makes it potentially more efficient than algorithmic methods.

Sections \ref{sec:description} and \ref{sec:methods} provide background on memristors and describe how they can be used to find the shortest path in a graph. Sections \ref{sec:control} and \ref{sec:ramp} present our method and validate it through numerical simulations of large numbers of randomly generated graphs of various sizes and topologies. We show that it is valid for realistic experimentally-derived memristor model and parameters, and that it is robust to device variability. Section \ref{sec:scaling} addresses the key advantage of this method by showing that the time and energy it consumes scales with the length of the shortest path. Finally, Section \ref{sec:implementation} addresses the hardware implementation of this method. 

\section{Memristor-based optimization \label{sec:description}}

Memristors are a class of devices that exhibit hysteretic behavior: their electrical conductance can be modified in a non-volatile fashion. The conductance can take values between two extreme states. The device is said to be "ON" or "OFF" when in its highest or lowest conductance state, respectively. The conductance can be repeatedly increased and decreased by running current through the device.\cite{strukov_missing_2008,pickett_switching_2009,menzel_origin_2011} 

A graph can be represented by a circuit of memristors: the nodes of the graph are electrical junctions connected by memristors implementing the edges of the graph. An example is shown in Figure \ref{fig:maze}(a): each black dot corresponds to a node and each colored rectangle to a memristor. The goal is to find the shortest path between the start and end nodes, marked by green stars in Figure \ref{fig:maze}(a). 

A voltage is applied across these nodes, causing current to flow through the circuit. The system is initialized with all memristors in their OFF state. The shortest path, i.e., composed of the lowest number of memristors, is more conductive than the other paths. As a consequence of Kirchhoff laws, more current flows through the memristors on the shortest path than through the other memristors. This causes their conductance to increase, drawing even more current to them and thus creating a reinforcement mechanism. On the other hand, the memristors outside the shortest path have little current flowing through them so their conductance does not increase. This behavior is shown in Figure \ref{fig:maze}(b) which presents the evolution of the conductance of the memristors belonging to the shortest path (in red) and the others (in black) versus time. After some time, the system reaches a steady state where the memristors on the shortest path are ON while the others are OFF. This state is depicted in Figure \ref{fig:maze}(a), where the conductance of each memristor is represented by the color of the rectangle. The successfully found shortest path can be visually observed in red, contrasting with the rest of the graph in purple. As their state is non-volatile, the voltage across the circuit can be turned off and the individual memristors can be measured in order to determine the shortest path (as described in Section \ref{sec:control}). 

Note that the shortest path is found here without supervision: it emerges from the dynamical evolution of the system. Furthermore, the computation is parallel, with each memristor evolving at the same time. This parallelism is key to how the time and energy consumed by the computation scale as the size of the problem increases, as detailed in Section \ref{sec:scaling}. 

\section{Memristor models \label{sec:methods}}
In this work we consider both a simple generic memristor model as well as a more complex realistic memristor model, derived from experiments.
We start by studying a simple generic memristor model, where
\begin{equation}
I = V \left( G_{\mathrm{ON}}x + G_{\mathrm{OFF}}\left(1-x\right) \right)
\end{equation}
where $I$ is the current though a memristor and $V$ the voltage across it. $G_{\mathrm{ON}}~=~10^{-1}~\mathrm{S}$ and $G_{\mathrm{OFF}}~=~10^{-4}~\mathrm{S}$ are the extreme conductance states of the memristors. $x$ is an internal state variable, bounded between 0 and 1, describing the conductance dynamics of the device. The conductance can be increased by running current through the device.
\begin{equation}
\frac{dx}{dt} = \gamma | I| - \frac{x}{\tau}
\end{equation}
where $\gamma = 10^6~\mathrm{A^{-1}s^{-1}}$, and $\tau~=~0.1~\mathrm{s}$ is the decay constant of the device.  The decay term shows that the devices are not strictly non-volatile, but are practically so on useful time scales.  We numerically simulate the evolution of circuits of such devices by solving Kirchhoff equations at each junction. Because of the linear relationship between current and voltage in this model, the dynamics of the circuit can be described as a system of linear equations, which allow us to perform a large number of simulations.

In order to validate our results, we also use the memristor model developed by Chang \textit{et al}., with the device parameters they extracted from experiments. \cite{chang_synaptic_2011} The model corresponds to a $\mathrm{Pd/WO_3/W}$ stack. Applying a voltage across the device induces oxygen vacancy migration in the oxide. The width and length of the created conducting filament determine the conductance of the device, through the internal state variable $x$.  
\begin{equation}
I = \left(1-x\right) \alpha \left(1-\exp\left(-\beta V\right)\right)+x \gamma \sinh\left(\delta V\right) 
\end{equation}
\begin{equation}
\frac{dx}{dt} = \lambda \left(\exp \left(\eta_1 V \right) - \exp \left(-\eta_2 V \right) \right)- \frac{x}{\tau}
\end{equation}
where $\alpha =~5~\times~10^{-7}~\mathrm{S}$, $\beta~=~0.5~\mathrm{V^{-1}}$, $\gamma~=~4~\times~10^{-6}~\mathrm{S}$, $\delta~=~2~\mathrm{V^{-1}}$, $\lambda~=~4.5~\mathrm{s^{-1}}$, $\eta_1~=~0.004~\mathrm{V^{-1}}$, $\eta_2~=~4~\mathrm{V^{-1}}$ and $\tau~=~10~\mathrm{s}$ are device parameters. 

Note that in this model, the polarity of the voltage matters, as it can grow or shrink the filament. Since most optimization problems map to directed graphs, devices sensitive to the polarity of the voltage could be an asset. For broader applications, schemes where two memristors of opposed polarities are connected in parallel for each edge could be used, as proposed in Ref.~\onlinecite{pershin_self-organization_2013}. Furthermore, other types of memristors, such as phase-change memories, are not sensitive to the polarity of the voltage across them.\cite{lankhorst_low-cost_2005,ielmini_phase_2011,lai_current_2003,bez_chalcogenide_2009} The non-linear form of the current-voltage relationship complicates the solution of the system of equations, leading us to use a commercial circuit simulator to solve the system of equations.


\section{Control voltage \label{sec:control}}

\begin{figure}
\centering
\includegraphics[width=3.5in]{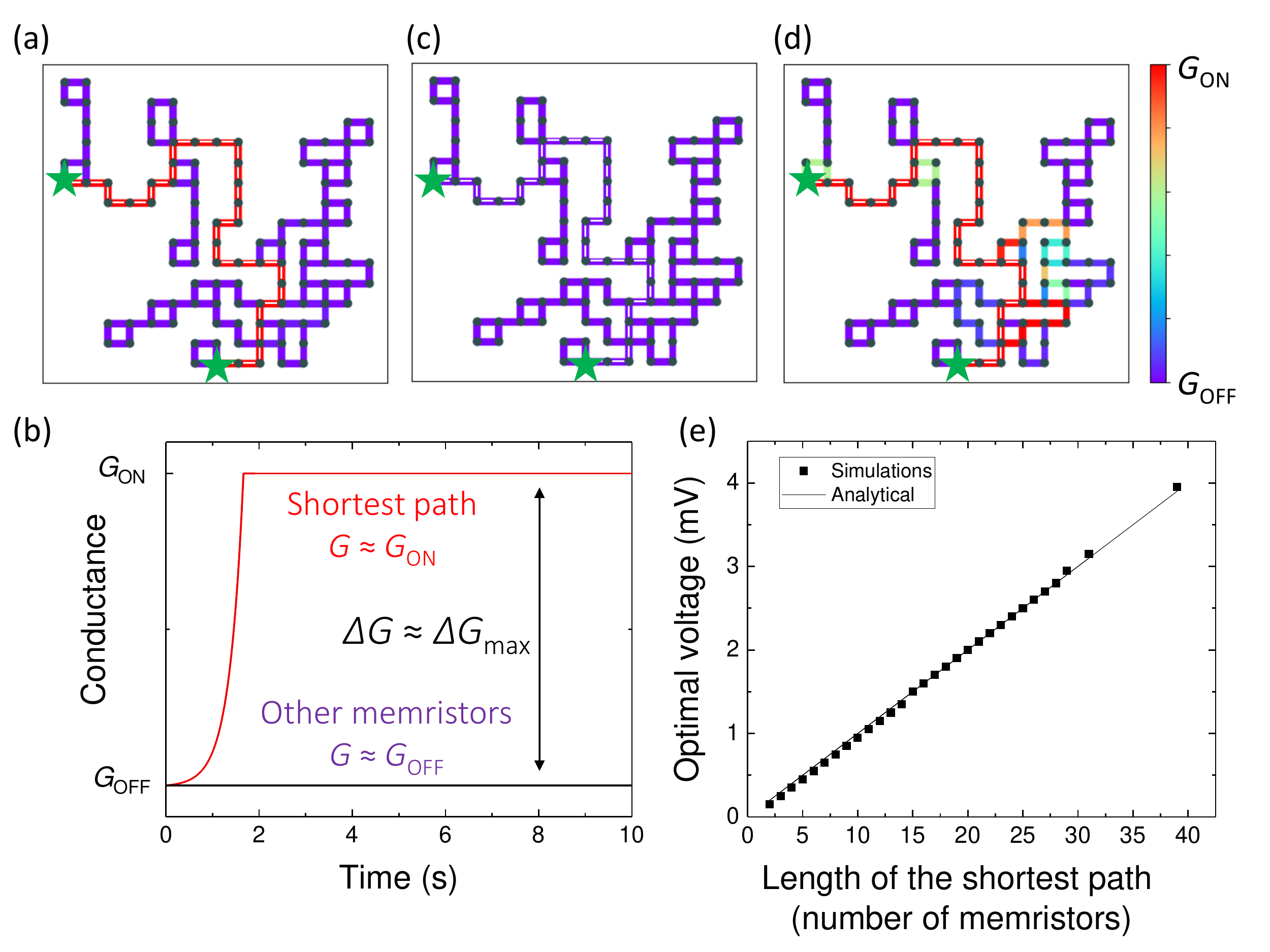}
\caption{(a)-(c)-(d) Schematic of a memristor circuit at its steady state for different applied voltages: 4~mV for panel (a), 1~mV for panel (c) and 10~mV for panel (d). The colors correspond to the conductance of each memristor. (b) Conductance versus time for the shortest path memristors (in red) and the longer paths memristors (in black). Note that for these ideal devices, the curves of all memristors in each path category superimpose. The difference $\Delta G$ is indicated by a double arrow. This corresponds to the graph depicted in panel (a), with an applied voltage of 4~mV. (e) Optimal voltage (leading to the highest $\Delta G$) versus the length of the shortest path, i.e., the number of memristors composing it. Each data point corresponds to a simulation on one of the 1996 randomly generated square grid graphs. The solid line correspond to the analytical value $\frac{N}{\gamma\tau G_{\mathrm{ON}}}$.
}
\label{fig:maze}
\end{figure}

We examine the importance of the control voltage, i.e., the voltage applied across the circuit, and show that using a constant voltage is not practical for applications. Panels (a), (c) and (d) of Figure \ref{fig:maze} show the steady state of one single graph after application of three different voltages, as described in Section \ref{sec:description}. The generic linear memristor model described in Section \ref{sec:methods} was used.
In Figure \ref{fig:maze}(a), where a voltage of 4 mV was applied, the computation was successful. However, Figures \ref{fig:maze}(c) and \ref{fig:maze}(d) show examples of incorrect computation. In Figure \ref{fig:maze}(c) the voltage (1 mV) was too low to significantly increase the conductance of any memristor. On the other hand, in Figure \ref{fig:maze}(d), the voltage (10 mV) was too high so the conductance of some memristors on longer paths was increased comparatively to those on the shortest path. Only an optimal range of control voltage leads to a successful computation of the shortest path.

To investigate the role of the control voltage on the computation, we define a metric of success for solving the shortest path problem. The input of the problem is the topology of the memristor circuit, i.e., the graph. In order to read the output of the computation (once steady state is reached or after the evolution of the system is stopped by the user, as we describe in Section \ref{sec:ramp}), we proceed as follows. The shortest path starts with the start node -- where the voltage source is applied -- and is constructed node by node. The conductance of each memristor connected to this node is measured. The next node on the shortest path is the one connected to the current node through the highest conductance memristor. This process is repeated until the end node -- the node connected the ground -- is reached. This method allows the user to read the shortest path by probing only a fraction of the memristors in the network, proportional to the length of the shortest path. For this method to read the correct shortest path, it is required that at each step, the measured memristor belonging to the shortest path has a higher conductance than the others. We thus define as metric of success, $\Delta G$, the smallest difference in conductance between a memristor on the shortest path and a memristor outside the path but connected to the same node of the path. The shortest path problem is successfully solved if $\Delta G > 0$. The highest possible success is $\Delta G _{\mathrm{max}} = G_{\mathrm{ON}} - G_{\mathrm{OFF}}$.
In the successful example of Figure \ref{fig:maze}(a), $\Delta G \simeq \Delta G _{\mathrm{max}}$, as observed in Figure \ref{fig:maze}(b). On the other hand, the results in Figures \ref{fig:maze} (c) and (d) both exhibit $\Delta G \simeq 0$.

In order to determine the optimal control voltage, we have generated thousands of graphs of different sizes, based on a square grid with randomly removed edges, similar to the one shown in Figure \ref{fig:maze}. The start and end nodes were selected randomly for each graph. Dead end branches, which would lead to floating memristors, where removed and so were nodes not connected to the start and end nodes. We only selected graphs that had a unique shortest path. We generated graphs using different grid sizes and different probabilities for edges to be removed. The correct shortest path was determined using the standard Breadth First Search algorithm.

For each graph, the corresponding circuit was numerically simulated at various control voltages. The  optimal control voltage is the voltage at which $\Delta G$ is the highest. Figure \ref{fig:maze}(e) shows that the optimal voltage is proportional to the length of the shortest path. In order to turn on a memristor, the time derivative of its $x$ must be non-negative when $x$ approaches 1, i.e, the voltage across the device is greater than $\frac{1}{\gamma \tau G_{\mathrm{ON}}}$. The lowest control voltage to achieve this is $\frac{N}{\gamma\tau G_{\mathrm{ON}}}$, where $N$ is the number of memristors on the shortest path. This optimal control voltage will turn on the shortest path but not any longer paths. As observed in Figure \ref{fig:maze}(e), this matches simulation results. 
Our simulations show that using a constant control voltage too far from the optimal control voltage gives results like those in panels (c) and (d) Fig.~\ref{fig:maze}, making such an appraoch impractical
for applications for which the length of the shortest path is not known in advance.  

\section{Using a voltage ramp \label{sec:ramp}}
\begin{figure}
\centering
\includegraphics[width=3.5in]{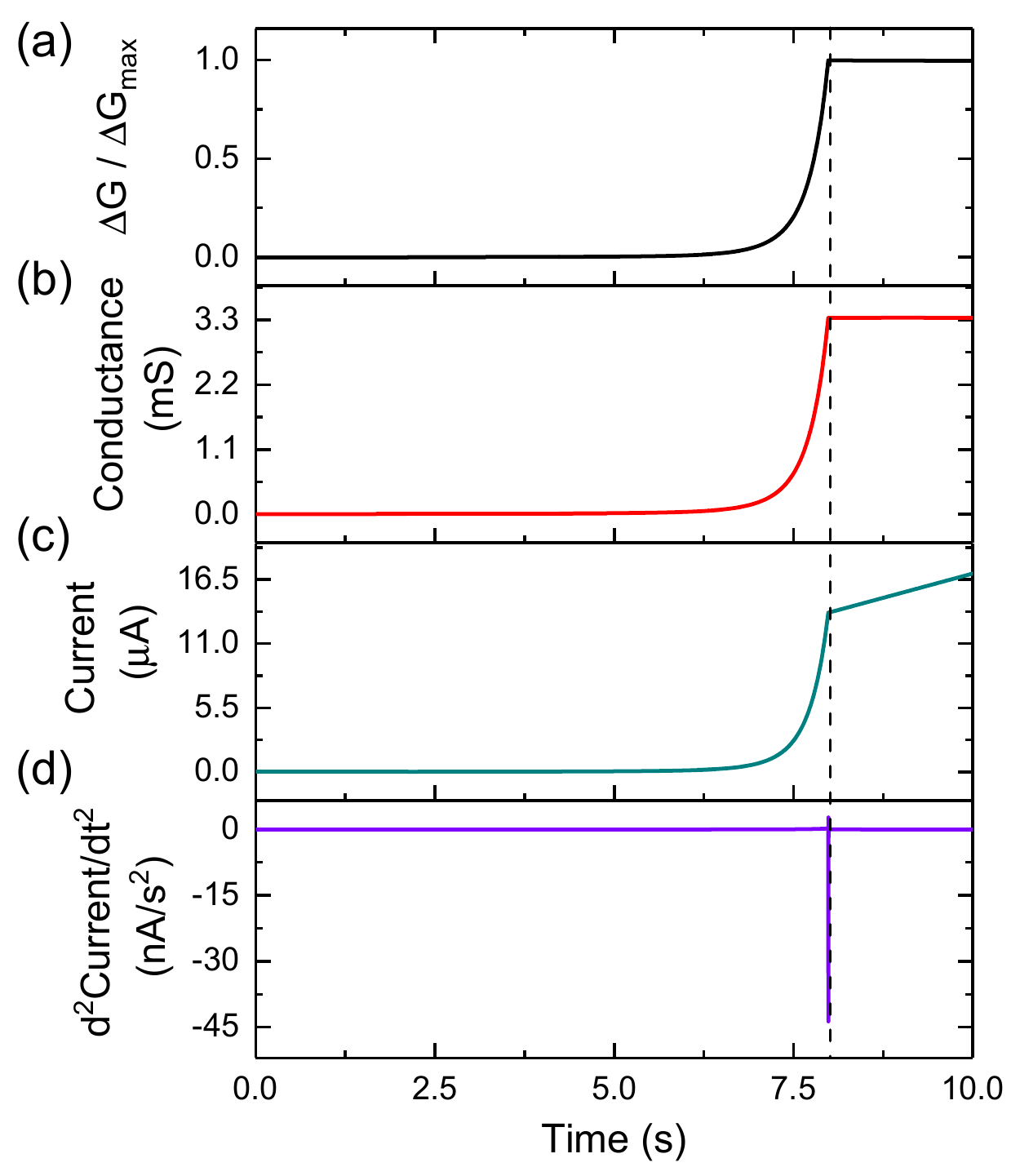}
\caption{Evolution with time of (a) the conductance difference $\Delta G / \Delta G _{\mathrm{max}}$, (b) the global conductance of the circuit, (c) the current through the circuit and (d) the second time derivative of this current. The voltage ramp starts from 0.1 mV and increases at a rate of $0.5~\mathrm{mV/s}$. The dashed line corresponds to the time of result detection.
}
\label{fig:ramp}
\end{figure}

We propose a method that does not require prior knowledge of the shortest path length. We leverage the fact that the shortest path is turned on (i.e., has the conductances of its memristors increase to $G_{\mathrm{ON}}$) at lower control voltages than longer paths, which suggest the use of a voltage ramp. Figure \ref{fig:ramp}(a) shows the evolution of $\Delta G$ with time as the voltage is increased. We observe a sharp increase in $\Delta G$, which corresponds to the shortest path turning on. The increase in conductance of the memristors on the shortest path creates an increase in the global conductance of the circuit, shown in Figure \ref{fig:ramp}(b). This increase can be detected by measuring the current going in and out of the circuit. The turning on of the shortest path corresponds to a sharp kink in the current, as shown in Figure \ref{fig:ramp}(c), and thus a drop below zero in the second time derivative of the current, as shown in Figure \ref{fig:ramp}(d). When this drop is measured, the control voltage is turned off and the result of the computation is read out as described in Section \ref{sec:control}. Note that here the evolution of the system is stopped, contrary to the constant control voltage method where a steady state is reached. 

\begin{figure}
\centering
\includegraphics[width=3.5in]{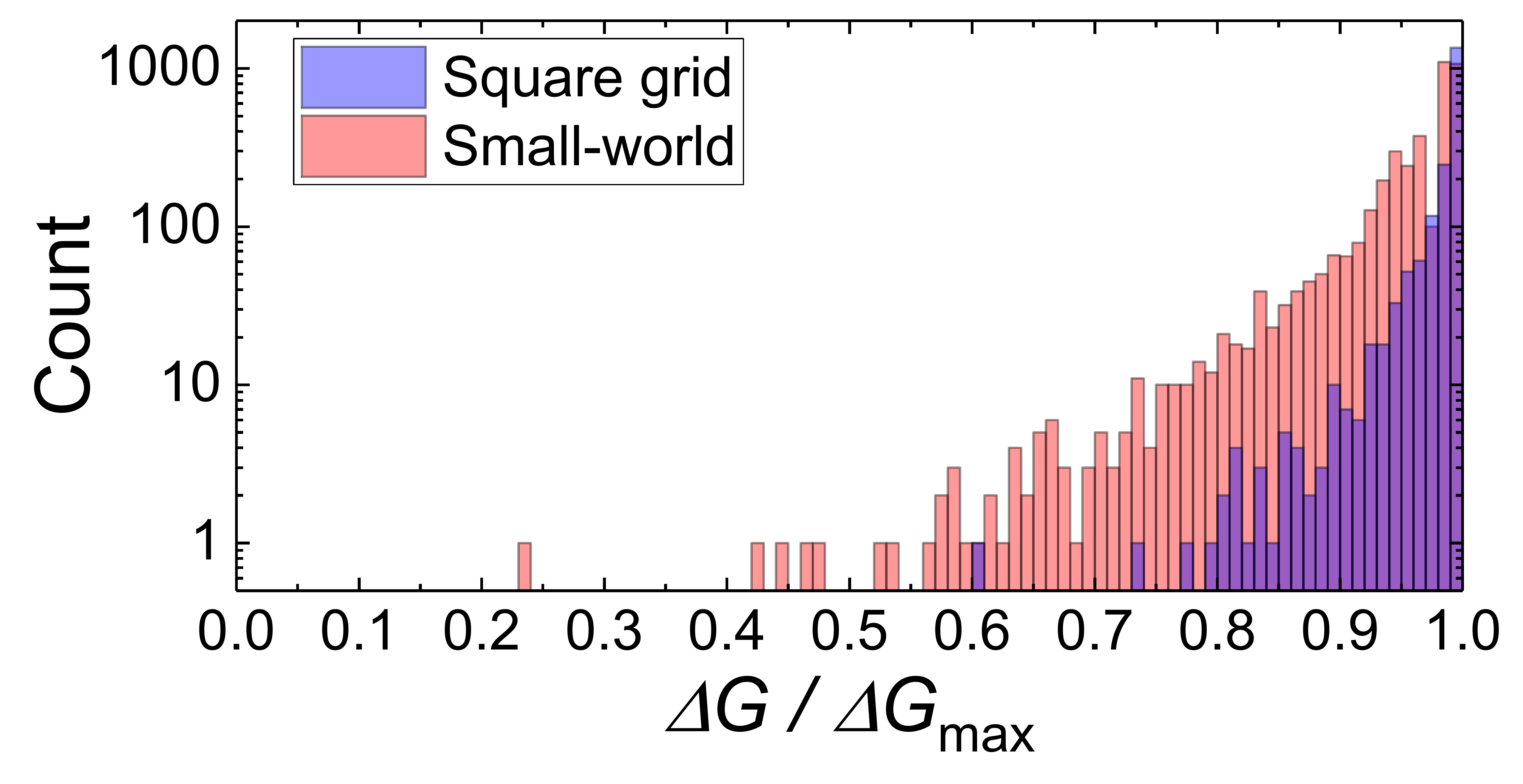}
\caption{Histograms of the metric of success $\Delta G / \Delta G _{\mathrm{max}}$ for simulations on randomly generated graphs, using the generic linear model on 1996 square grid based graphs (blue bars) and 4797 small-world networks (red bars). The voltage ramp starts from 0.1 mV and increases at a rate of $0.5~\mathrm{mV/s}$.
}
\label{fig:perf}
\end{figure}

Figure \ref{fig:perf} presents statistics on $\Delta G$ obtained from simulations and shows that this method, using a single voltage ramp, can successfully find the shortest path in thousands of graphs of various sizes. In addition to the square grid based topology (blue bars), we have performed similar simulations on graphs with a small-world topology (red bars). Small-world networks stand in-between regular and random graphs and describe many interesting problems\cite{watts_collective_1998} and are discussed in more detail in Sec.~\ref{sec:scaling}. Here we have generated small-world networks of different sizes and levels of randomness. For both topologies, all simulations exhibit $\Delta G$ well above zero, which validates this method. 

\begin{figure}
\centering
\includegraphics[width=3.5in]{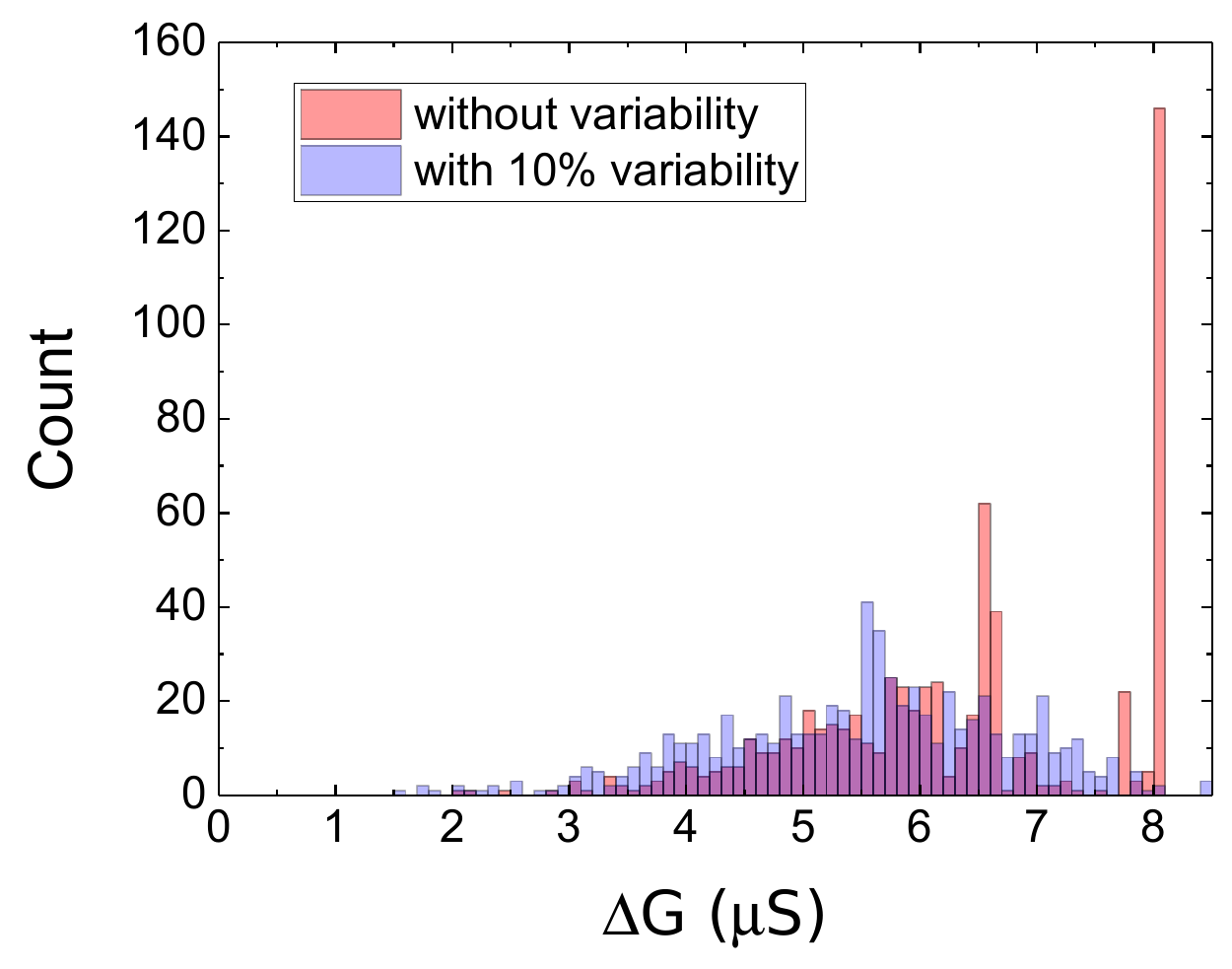}
\caption{Histograms of the metric of success $\Delta G$ for simulations on 658 randomly generated graphs following the square grid topology, using the experimentally obtained realistic model from Ref. \onlinecite{chang_synaptic_2011} without variability (red bars) and with $10~\%$ variability on all device parameters (blue bars). With device variability, each simulation is done on one graph with one set of parameters randomly chosen from a Gaussian distribution around the nominal parameter value and of $10~\%$ standard deviation. The set of graphs with and without variability are the same. The voltage ramp starts from 0 V and increases at a rate of $1~\mathrm{mV/s}$.
}
\label{fig:perf_real}
\end{figure}

We test the validity of our method for realistic devices by performing simulations on a square grid topology with the realistic memristor model and parameters described in Section \ref{sec:methods}.\cite{chang_synaptic_2011} Figure \ref{fig:perf_real} shows that $\Delta G$ is more widely spread than for the generic linear model. This is due to the fact that the realistic model produces smoother behavior and a weaker reinforcement mechanism. However, our approach remains valid, as all graphs exhibit $\Delta G >0$. 

Furthermore, we investigate the influence of device variability. As shown in Figure \ref{fig:perf_real}, even with $10~\%$ variability on all device parameters, our approach still works. Interestingly, variability can even be an asset. For simplicity, in this study, we have restricted ourselves to shortest path problems with a unique solution. However, we have observed that, in the case of graphs in which there are two shortest paths of equal length between the considered nodes, systems with ideal devices tend to have lower $\Delta G$ than in the case with unique solutions. Qualitatively, the system tries to turn all shortest paths on simultaneously, which prevents the winner-take-all reinforcement mechanism to take place properly. However, device variability makes one of the shortest paths intrinsically more conductive and easier to turn on, which lets the system choose this path over the others and turn it on completely, thus, increasing $\Delta G$. This is an interesting example of how, in bio-inspired computing, features of nanodevices usually seen as drawbacks can be beneficial. 

\section{Scaling of the time and energy consumption \label{sec:scaling}}

In order to evaluate the potential use of our method, we study how the time and energy required by the computation scale with the size of the graph. The energy was estimated as the integral over time of the total current through the circuit times the voltage across the circuit. Additional energy will be spent for detecting the current second time derivative, reading the result and setting up the circuit, but this is out of the scope of this study as designing the full architecture of the system would be required. 

\begin{figure}
\centering
\includegraphics[width=3.5in]{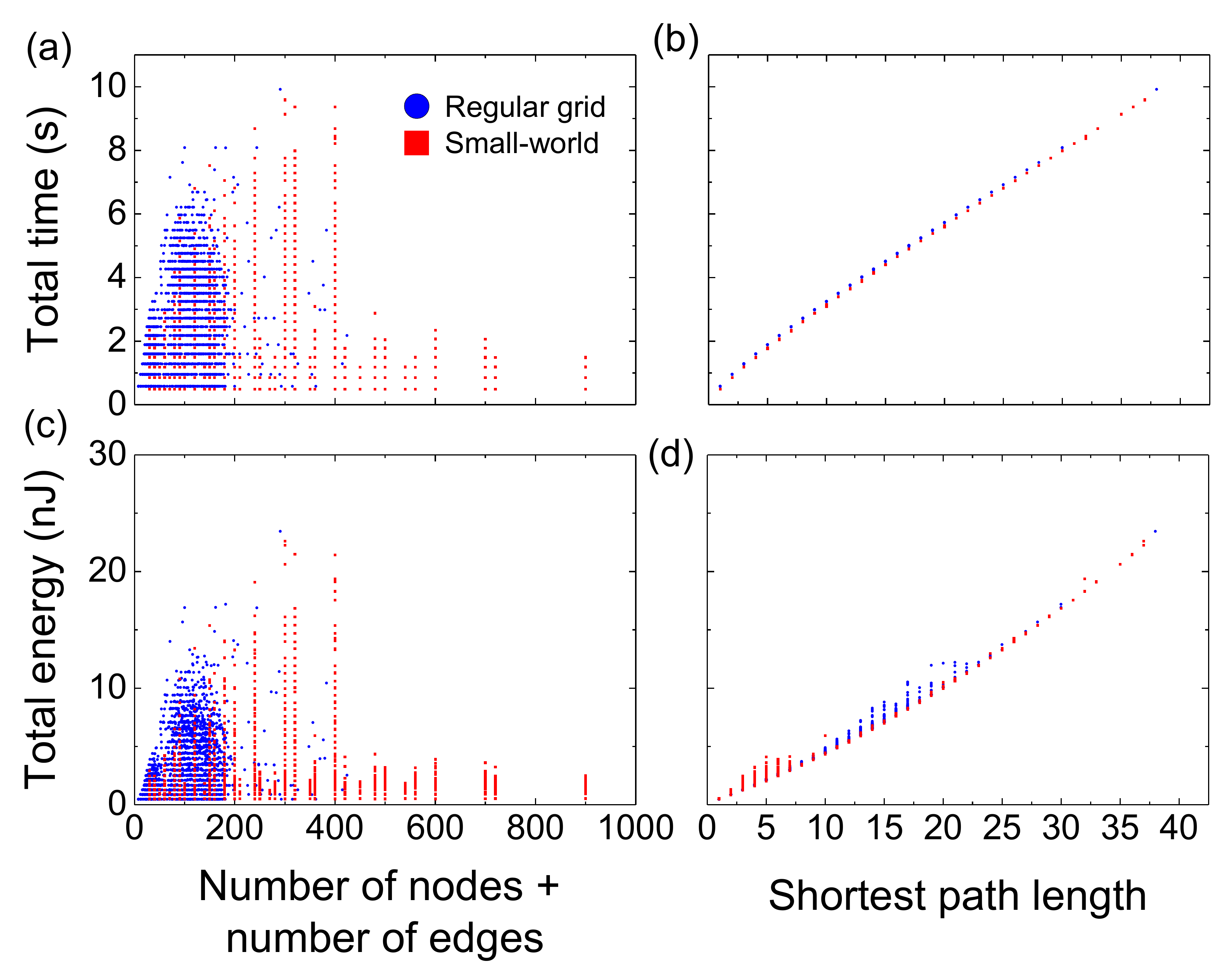}
\caption{Total time of the computation versus (a) the size (number of nodes plus number of number of edges) and (b) the length of the shortest path. Total energy consumed versus (c) the size and (b) the length of the shortest path. Blue circles correspond to square grids and red squares correspond to small-world networks. Each symbol corresponds to one of the 1996 square grid and 4797 small-world graphs.
}
\label{fig:scaling}
\end{figure}

Figures \ref{fig:scaling}(a) and \ref{fig:scaling}(b) present the time and energy consumption versus the number of nodes plus the number of edges for all our simulations with the generic model (corresponding to Figure \ref{fig:perf}). This combination is a common way to characterize the size of a graph.\cite{dijkstra_note_1959}  We observe no correlation. However, Figures \ref{fig:scaling}(c) and \ref{fig:scaling}(d) show that the time and energy consumption correlate strongly with the length of the shortest path. Moreover, this scaling does not depend on the topology of the graph. 

These results indicate a key advantage of the proposed method. Conventional algorithmic methods typically scale with the number of nodes and the number of edges, because the different nodes and edges are explored sequentially.\cite{dijkstra_note_1959,ahuja_faster_1990,cherkassky_shortest_1996,thorup_undirected_1999,thorup_integer_2004,fredman_fibonacci_1984,williams_faster_2014} In the present hardware implementation, the current explores the entire circuit in parallel, which makes the time and energy consumption independent of the size of the graph. This method would be particularly efficient for large graphs with small shortest paths. Such graphs include small-world networks. Standing between regularity and randomness, these are composed of many short range connections and a few long range connections. They exhibit high clustering and low shortest paths.\cite{watts_collective_1998} Small-world networks have been shown to describe many systems with important applications, such as power grids, the structure of the web, social media, and neural networks.\cite{watts_collective_1998,albert_internet:_1999,bassett_small-world_2006,palla_uncovering_2005}

\begin{figure}
\centering
\includegraphics[width=3.5in]{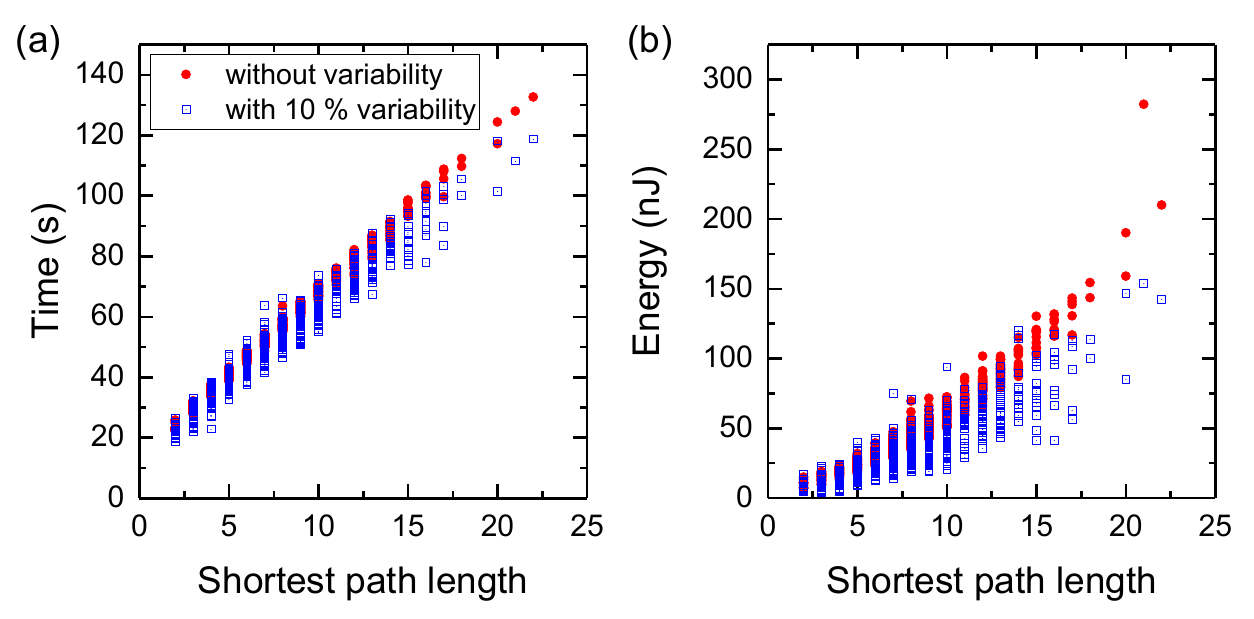}
\caption{(a) Total time of the computation versus the length of the shortest path. (b) Total energy consumed versus the length of the shortest path. Each symbol corresponds to a simulation using the realistic memristor model either without variability (red full circles) or with $10~\%$ variability on each parameter (blue squares), on one of the 658 square grid graphs. With device variability, each simulation is done on one graph with one set of parameters randomly chosen from a Gaussian distribution around the nominal parameter value and of $10~\%$ standard deviation. The set of graphs with and without variability are the same. 
}
\label{fig:scaling_real}
\end{figure}

We investigate the effect of a realistic memristor model and device variability by computing the time and energy consumption corresponding to the simulations used in Figure \ref{fig:perf_real}. We observe that the scaling laws stay valid: as shown in Figure \ref{fig:scaling_real}, the time and energy consumption depend on the length of the shortest path. The fact that the time and energy consumption appear to scale independently of the topology or memristor model is promising for the ease of material implementation, as many types of devices could be used, and for the breadth of applications, as many types of graphs could be solved.

\section{Outlook for a reconfigurable graph solver \label{sec:implementation}}
In order for this approach to be of practical use, it must be reconfigurable for different problems. Finding the shortest path between different nodes of the same graph simply requires connecting the voltage source and ground to the new nodes. Modifying the graph is non-trivial. Nearest neighbor connections could be implemented by complementary metal-oxide-semiconductor (CMOS) switches that are opened or closed to form the desired graph, as proposed in Ref. \onlinecite{pershin_solving_2011}. Longer range connections could be implemented by another set of memristive devices, arranged in conventional crossbar topologies. The hybrid CMOS/molecular (CMOL) architecture, proposed in Ref. \onlinecite{strukov_cmol_2005}, has been shown to give high densities of connections, which could be useful here.\cite{strukov_monolithically_2010,madhavan_high-throughput_2018} Building one physical system capable of implementing any graph problem is unrealistic. However, it would be possible to have specialized chips for types of graph problems. There would be one underlying hierarchical structure appropriate to the graph type, as well as many reconfigurable connections implementing specific problems. Such hierarchical architectures have been shown to be efficient at simulating large artificial neural networks.\cite{wang_fpga-based_2018}

\section{Conclusion}
We propose a scheme to find the shortest path in a graph problem using a circuit of memristors. It includes procedures to detect and read the result. Our scheme does not require prior knowledge about the shortest path and has been validated on large number of graphs of various sizes and topology. We have shown that this scheme works for realistic device models and is robust to variability. It scales with the length of the shortest path in terms of time and energy consumption because of its intrinsic parallelism. This is a key advantage compared to conventional algorithmic methods that scale with the size of the graph. In particular, it would be best for studying large graphs with small shortest paths, such as social networks, neural networks, or power grids. These results are promising for hardware implementations of systems capable of performing fast and energy efficient analysis of large graphs. 

More broadly, the field of swarm intelligence is rich, and implementing its concepts in hardware offers many paths towards energy efficient computing. For example, it was shown that implementing a swarm intelligence algorithm of image edge detection with circuits of memristors consumes less energy than conventional methods.\cite{pajouhi_image_2018} Exploring other swarm intelligence ideas and different substrates to implement them is an exciting road towards low energy cost systems that perform complex optimization tasks. 

\begin{acknowledgments}
The authors acknowledge A. Madhavan, M. Daniels, N. Zhitenev, J. McClelland, R. McMichael, S. Dushenko and P. Shrestha  for helpful comments and discussions.
A.M. acknowledges support under the Cooperative Research Agreement between the University of Maryland and the National Institute of Standards and Technology, Center for Nanoscale Science and Technology, Grant No. 70NANB10H193, through the University of Maryland. This material is based upon work supported by the National Institute of Standards and Technology Summer Undergraduate Research Fellowship (SURF) Program under Grant No. 70NANB18H080.
\end{acknowledgments}


%

\end{document}